\documentclass[12pt]{article}

\usepackage{amsmath,amssymb,pictex,cite,eepic,epsfig,psfrag,url}
\advance \topmargin by -\headheight
\advance \topmargin by -\headsep
\evensidemargin \oddsidemargin
\makeatletter
\@addtoreset{equation}{section}

\makeatother

\makeatletter
\def\Appendix{\appendix
  \def\@seccntformat##1{Appendix~\csname the##1\endcsname.~~}}
\makeatother

\begin{document}
\hfill \hbox{FIAN-TD-01-11}
\vspace{1.5cm}



\bigskip

\begin{center}
{\Large \textbf{AGT conjecture and Integrable structure\\
of Conformal field theory for $c=1$}}

\vspace{1.5cm}

{\large A.~Belavin}

\vspace{0.2cm}

Landau Institute for Theoretical Physics, RAS, Chernogolovka, Russia

\vspace{0.2cm}

and

\vspace{0.2cm}

{\large V.~Belavin}

\vspace{0.2cm}
Theory Department, Lebedev Physical Institute, RAS, Moscow, Russia

\end{center}

\vspace{1.0cm}

\textbf{Abstract}

AGT correspondence gives an explicit expressions for the conformal blocks
of $d=2$ conformal field theory. Recently an explanation of this representation
inside the CFT framework was given through the assumption about the existence of the special orthogonal basis in the module of algebra $\mathcal{A}=Vir\otimes\mathcal{H}$.
The basis vectors are the eigenvectors of the infinite set of commuting integrals of motion. 
It was also proven that some of these vectors take form of Jack polynomials.
In this note we conjecture and verify by explicit computations that in the case of the Virasoro central charge $c=1$ all basis vectors are just the products of two Jack polynomials. Each of the commuting integrals of motion
becomes the sum of two integrals of motion of two noninteracting Calogero models.
We also show that in the case $c\neq1$ 
it is necessary to use two different Feigin-Fuks bosonizations of the Virasoro algebra for the construction of all 
basis  vectors which take form of one Jack polynomial.

\section{Introduction}

AGT conjecture~\cite{Alday:2009aq} reveals a deep connection between 2d CFT and $N=2$ SUSY gauge theories.
This correspondence turns out to be very important for the 2d CFT.
In particular, it gives a remarkable explicit representation for the conformal block coefficients in terms of the Nekrasov partition functions~\cite{Nekrasov}.
 This representation was not known in the framework of 2d CFT.
Certainly, it is a challenge for CFT. It is necessary to achieve appropriate understanding of this new representation for conformal blocks staying inside CFT frame. 
Different attempts to derive the Nekrasov representation for the conformal blocks using the conformal bootstrap approach~\cite{BPZ,AlZ} have been performed recently~\cite{Rubik,Leshek,FL}.
 Many new features, as well as some new connections between Seiberg-Witten theory, Matrix Models, Dotsenko-Fateev representation~\cite{DF} for conformal blocks, Selberg integrals, {\it etc}, were found (see e.g.~\cite{DV,EM,IO,MMS}). 

Recently an important for the understanding of the new representation for conformal blocks step was made in~\cite{AT,F,ALTF}.
 It was assumed by Alday and Tachikawa in~\cite{AT} that in the Hilbert space $\mathcal{H}_a$, 
formed by the tensor product of the Fock module, which corresponds to the so called $U(1)$ factor~\cite{Alday:2009aq}, and the 
Virasoro module $V_a$ a special orthogonal basis which consists of $|\vec{Y}\rangle$ for all pairs $\vec{Y}$ 
of Young diagrams can be constructed.
It was also assumed in~\cite{AT}, that
the matrix elements of the product of Carlsson and Okounkov vertex operator~\cite{CO}
and the Virasoro primary field between two vectors from two different copies of $\mathcal{H}_a$ 
have a very simple factorized form. These matrix elements coincide with the Nekrasov partition functions for the 
bifundamental hypermultiplet.
In~\cite{ALTF} some evidence of this proposal based on the explicit computations was given. 
It was also proven in~\cite{ALTF} that if first of two Young diagrams is empty the corresponding basis vector has 
form of Jack polynomial~\cite{Jack}. To get the Jack representation for such vectors it is necessary to use 
the Feigin-Fuks bosonization of the Virasoro algebra\footnote{For other connection between the Jack symmetric 
polynomials and the Virasoro algebra see~\cite{Shiraishi}.}.
It was assumed and verified in~\cite{ALTF} that this basis diagonalizes an infinite set of commuting Integrals of 
Motion of some integrable hierarchy.
However, the explicit form of the vectors of the orthogonal basis in the general case remains still unknown.

In this paper we continue to study this basis and the underlying integrable 
structure.
First of all we want to understand the form of the vectors for both particular cases 
when one of Young diagrams, the first or the second one, is empty.
In addition to the basis vectors $|Y,\varnothing \rangle$ constructed in~\cite{ALTF}
we construct the vectors of the form $|\varnothing,Y\rangle$.
 Their form is rather complicated in terms of the Feigin-Fuks bosonization used
 for constructing vectors $|Y,\varnothing\rangle$. However, once the second possible bosonization is used,  
 they simplify and also take the form of Jack polynomials.
The naive assumption that if both Young diagrams are not empty the vectors are 
the products of two Jack polynomials which depend on two different Feigin-Fuks bosons 
is not correct. 
However it is true in the case $c=1$. In this case the bosons of two different sets commute if their indices are 
of the same sign.  
We verified this statement by explicit computations.
This facts lead us to the conjecture that the Integrals of motion of~\cite{ALTF} for $c=1$ 
coincide with the sum of Integrals of Motion of two noninteracting Calogero models.
This statement was also verified using the explicit expressions given in~\cite{ALTF} for the first 
Integrals of motion.

The paper is organized as follows. In section $2$ we recall the definition of the 
conformal block functions of 2d CFT and  describe their relation to Nekrasov's partition functions.
In section $3$ we discuss the two possible versions of Feigin-Fuks bosonization of the Virasoro algebra and the notion of Liouville reflection operator~\cite{ZZ}.
This is important for the correct choice  of the arguments of Jack functions in some special cases,as well as in the formulation of the conjecture
about the form of basis vectors for $c=1$. 
In section $4$ we focus our attention on the case of central charge $c=1$. The Jack functions 
are reduced in this case to the Schur polynomials.
We conjecture that in this case all basis vectors are products of two Schur polynomials 
and verify this statement up to the level $3$.
 Some formulas related to the Nekrasov representation of the conformal block, the definition of the Jack symmetric functions and their basic properties 
are collected in the appendices.

\section{AGT correspondence and a special basis of states in
the highest weight representations of $Vir\otimes \mathcal{H}$}

Conformal blocks~\cite{BPZ} are special analytic functions on a
Riemann surface of genus $g$ with $n$ punctures. They play an
important role in $d=2$ CFT. The AGT conjecture~\cite{Alday:2009aq} provides us
with explicit expressions for the conformal blocks in terms of
the Nekrasov partition functions of a certain class of $N=2$
SCFTs~\cite{Gaiotto}. Below we shall consider the case of the $4$-point
conformal block on the sphere. In addition to the four-point projective invariant $q$
it depends on six parameters: the central charge $c$, four external conformal dimensions $\Delta_i$
and the dimension of the intermediate primary filed $\Phi_{\Delta}$ (for details see Appendix A).
It can be represented by the following diagram
\begin{equation}\label{conformal-block}
    \begin{picture}(-30,75)(100,10)
    \Thicklines
    \unitlength 2.3pt
    \put(0,0){\line(1,0){70}}
    \put(20,0){\line(0,1){25}}
    \put(50,0){\line(0,1){25}}
    \put(-7,-1){\mbox{$\Delta_{1}$}}
    \put(18,28){\mbox{$\Delta_{2}$}}
    \put(48,28){\mbox{$\Delta_{3}$}}
    \put(32,3){\mbox{$\Delta$}}
    \put(72,-1){\mbox{$\Delta_{4}$}}
    \end{picture}
    \vspace*{1cm}
\end{equation}
which defines one of three possible channels of fusing external fields into intermediate one.
Instead of the conformal dimensions $\Delta_i$, $\Delta$, we use sometimes the 
parameters $\lambda_i$, $P$: 
\begin{equation}
\Delta_i=Q^2/4-\lambda_i^2, \qquad \Delta=Q^2/4-P^2,
\end{equation}
while, instead of Virasoro central charge we introduce the parameter $b$:
\begin{equation}
c=1+6 Q^2, \qquad Q=1/b+b.
\end{equation}
In terms of new variables the $4-$point conformal block is denoted as
$F^{\text{\sf V}}(q,\lambda_i,P)$.
Due to the operator product expansion it has the form of a power series 
\begin{equation}
F^{\text{\sf V}}(q,\lambda_i,P)=\sum_{N=0}^{\infty}\, 
{}^{\text{\sf V}}\langle N;\lambda_1,\lambda_2|N;\lambda_3,\lambda_4\rangle^{\text{\sf V}}\,q^N,
\end{equation}
where $ |N;\lambda_1,\lambda_2\rangle^{\text{\sf V}}$ stands for the so-called chain vector,
which is defined as a linear combination of $N$th level descendants $L_{-k_1}L_{-k_2}\dots|P\rangle$ 
($L_n$ being the generators of the Virasoro algebra).
Conformal symmetry determines uniquely this
function\cite{BPZ}. It implies that
\begin{equation}
L_n |N;\lambda_1,\lambda_2\rangle^{\text{\sf V}} 
=(\Delta+n \Delta_1-\Delta_2+N-n) |N-n;\lambda_1,\lambda_2\rangle^{\text{\sf V}},
\label{Vir-chain}
\end{equation}
for any positive $n$. To formulate the AGT conjecture 
we introduce function 
\begin{equation}
F(q)=F^{\mathcal H}(q)F^{\text{\sf V}}(q),
\end{equation}
where $F^{\mathcal H}(q)$\footnote{To simplify notation,
we will suppress sometimes some of the arguments of the conformal blocks.} stands for
the Heisenberg conformal block 
\begin{equation}
F^{\mathcal H}(q)=(1-q)^{2(\frac{Q}{2}+\lambda_1)(\frac{Q}{2}-\lambda_3)}.
\end{equation}
The term ``Heisenberg'' for $F^{\mathcal H}(q)$ is related to the following
interpretation of this function. Let $a_n$ be generators of the Heisenberg algebra
$\mathcal{H}$ with the commutation relations
\begin{equation}
[a_n,a_m]=\frac{n}{2}\delta_{n+m,0}.
\end{equation}
The vectors $|N,\alpha\rangle^{\mathcal H}$ in the module $\mathcal{H}$ with the highest vector
$|0\rangle$ (i.e. $a_n|0\rangle=0$ for $n>0$),
is defined as the only linear
combination of vectors $|\vec{k}\rangle=a_{-1}^{k_1} a_{-2}^{k_2}\dots|P\rangle$ on the level $N$ that 
obeys the recursive relation
\begin{equation}
a_n |N,\alpha\rangle^{\mathcal H} =\alpha |N-n,\alpha\rangle^{\mathcal H}
\label{H-chain}
\end{equation}
One can evaluate explicitly 
\begin{equation}
|N,\alpha\rangle^{\mathcal H}=
\sum_{k_1,k_2,\dots} \prod_{l=1}^{\infty} \frac{(2\alpha)^{k_l}}{k_l!\, l^{k_l}}|\vec{k}\rangle
\end{equation}
Once the conjugation $(a_k)^+=a_{-k}$ is defined, the Heisenberg conformal block
is expressed in terms of the vectors $ |N,\alpha\rangle^{\mathcal{H}}$  
\begin{equation}
F^{\mathcal H}(q)=\sum_N \,^{\mathcal{H}}\langle N,i (\frac{Q}{2}-\lambda_3)
|N,i (\frac{Q}{2}+\lambda_1)\rangle^{\mathcal{H}}\,q^N.
\end{equation}
Now one can build the ``mixed'' block $F(q)$ from the vectors $|N,\lambda_1,\lambda_2\rangle$,
which belong to the $\mathcal A$-module, the representation
space of the algebra $\mathcal A=Vir \otimes \mathcal{H}$.
(The $\mathcal A$-module is just the tensor product 
of the $\mathcal{H}$-- and of the Virasoro--modules)
\begin{equation}
F(q)=\sum_N \langle N,\lambda_3,\lambda_4 |N,\lambda_1,\lambda_2\rangle\,q^N,
\end{equation}
where 
\begin{equation}
|N,\lambda_1,\lambda_2\rangle=
\sum_{\substack{N_1,N_2\\N_1+N_2=N}}|N_1,i\lambda_1\rangle^{\mathcal H} |N_2;\lambda_1,\lambda_2\rangle^{\text{\sf V}}.
\label{mixed-chain-vector}
\end{equation}
In these notation the AGT conjecture claims that
\begin{equation}
\langle N,\lambda_3,\lambda_4|N,\lambda_1,\lambda_2\rangle=
\sum_{\substack{Y_1,Y_2\\|Y_1|+|Y_2|=N}}\frac{Z_f(\vec{a},\vec{Y},\mu_1) Z_f(\vec{a},\vec{Y},\mu_2)
 Z_{af}(\vec{a},\vec{Y},\mu_3) Z_{af}(\vec{a},\vec{Y},\mu_4)}{D(\vec{a},\vec{Y})\bar{D}(\vec{a},\vec{Y})}.
\label{mixedscalprod}
\end{equation}
Here, in the right hand side the summation runs over pairs of Young tableaux $(Y_1,Y_2)$
with the total number of cells equal to $N$. The explicit form of
$Z_f(\vec{a},{\vec Y},\mu)$, $Z_{af}(\vec{a},{\vec Y},\mu)$ and $D(\vec{a},\vec{Y})$, 
$\bar{D}(\vec{a},\vec{Y})$ can be found in Appendix A. 
The parameters of Nekrasov's partition function
are related to the parameters of the conformal block as follows:
\begin{equation}
\begin{aligned}[centered]
\mu_1=\frac Q2-(\lambda_1+\lambda_2),\qquad \mu_2=\frac Q2-(\lambda_1-\lambda_2), \\
\mu_3=\frac Q2-(\lambda_3+\lambda_4),\qquad \mu_4=\frac Q2-(\lambda_3-\lambda_4), \\
\end{aligned}
\label{parameters-mu}
\end{equation}
and 
\begin{equation}
\vec{a}=(a,-a),\qquad a=P.
\label{parameter-a}
\end{equation}
The form of the scalar product~\eqref{mixedscalprod}
as a sum of simple rational Nekrasov functions
leads to the natural idea \cite{F,ALTF} that the vector $|N\rangle$ can be written as a linear combination of
some orthogonal vectors $|Y_1,Y_2\rangle$ which form a basis in the $\mathcal A$-module.
To ensure~\eqref{mixedscalprod}, 
the expansion of $|N\rangle$ over this basis should be 
\begin{equation}
|N,\lambda_1,\lambda_2\rangle=
\sum_{\vec Y, |\vec Y|=N} \frac{Z_f(\vec{a},\vec{Y},\mu_1) Z_f(\vec{a},\vec{Y},\mu_2)}{D(\vec{a},\vec{Y})}|\vec Y\rangle.
\end{equation}
The basis vectors up to the level $6$ and of the form $|Y_1,\varnothing\rangle$ on the general level were calculated in \cite{ALTF}. When one of Young tableaux 
is empty the corresponding vector is expressed through the Jack polynomials.\\
To write down the known basis vectors \cite{ALTF} explicitly,
 as well as to formulate the conjecture
about the form of vectors $|Y_1,Y_2\rangle$ in the special case $c=1$, 
it is necessary to use the Feigin-Fuks representation for the Virasoro modules.
It will be done in the next section.

\section{Two bosonizations of Virasoro algebra and
Liouville reflection operator}
Let us consider the Heisenberg algebra with a set of generators $b_k$ and $\hat P$, commuting
as
\begin{equation}
[b_n,b_m]=\frac{n}{2}\delta_{n+m,0},\qquad [b_n,\hat{P}]=0.
\end{equation}
The Feigin-Fuks representation~\cite{FF} for Virasoro algebra is
\begin{equation}
\begin{aligned}
&L_n=\sum_{k \neq 0,n}b_k b_{n-k}+i(nQ+2\hat{P})b_n,\\
&L_0=2\sum_{k>0}b_{-k}b_k+\frac{Q^2}{4}-{\hat P}^2.
\end{aligned}
\end{equation}
Fock space with the vacuum vector $|P\rangle$, such that $b_k|P\rangle=0$ for $k>0$ and $\hat{P}|P\rangle=P|P\rangle$,
is the Virasoro highest weight representation with the conformal dimension $\Delta=\frac{Q^2}{4}+P^2$ and
the central charge $c=1+6Q^2$.
A second set of the generators, which we denote $b_k^{\text{R}}$, exist in the universal  
enveloping Heisenberg algebra.
It is connected with the previous one by some unitary transform $\hat{S}(P)$
\begin{equation}
b_k^{\text{R}}=\hat{S}(P)\,b_k\,\hat{S}^{-1}(P).
\end{equation}
The requirement, which fixes the new generators $b_k^{\text{R}}$,
is that they should give the second possible Feigin-Fuks representation of the
Virasoro algebra
\begin{equation}
L_n=\sum_{k \neq 0}\, b_k^{\text{R}} b_{n-k}^{\text{R}}+i(nQ-2\hat{P})b_n^{\text{R}}.
\label{second-bosonization}
\end{equation}
The operator  $\hat{S}(P)$ is called reflection operator~\cite{ZZ}.
It plays some special role in Liouville field theory.
In general, this is some nonlinear transformation which is unknown in the closed form.
However, the first terms of the formal expansion can be constructed explicitly. 
Say, for $b_{-1}^{\text{R}}$, $b_{-2}^{\text{R}}$,
we have the following system of the constraints
\begin{equation}
\begin{aligned}
L_{-1}(b_k^{\text{R}},-\hat{P})|P\rangle=L_{-1}(b_k,\hat{P})|P\rangle,\\
L_{-1}^2(b_k^{\text{R}},-\hat{P})|P\rangle=L_{-1}^2(b_k,\hat{P})|P\rangle,\\
L_{-2}(b_k^{\text{R}},-\hat{P})|P\rangle=L_{-2}(b_k,\hat{P})|P\rangle,\\
\end{aligned}
\end{equation}
which should be considered as the equations for unknown $b_{-1}^{\text{R}}$ and $b_{-2}^{\text{R}}$.
Then the  first terms of the series for $b_{-1}^{\text{R}}$, $b_{-2}^{\text{R}}$ are
\begin{equation}
\begin{aligned}
b_{-1}^{\text{R}}=
\frac{Q-2p}{Q+2p}b_{-1}+&\frac{4iPQ}{(2P-Q)(1+4P^2+6PQ+2Q^2)}b_{-2}b_1-\\
&-\frac{16PQ}{(2P+Q)(2P-Q)(1+4P^2+6PQ+2Q^2)}b_{-1}^2b_1+\dots\,,\\
\end{aligned}
\end{equation}
\begin{equation}
\begin{aligned}
&b_{-2}^{\text{R}}=-\frac{2P+8P^3-Q-6PQ^2-2Q^3}{(2P+Q)(1+4P^2+6PQ+2Q^2)}b_{-2}+
\frac{8iPQ}{(2P+Q)(1+4P^2+6PQ+2Q^2)}b^2_{-1}+\dots\,.
\end{aligned}
\end{equation}
The standard form of the Virasoro conjugation $L_n^{+}=L_{-n}$ fixes the conjugation
of the bosonizing Heisenberg generators
\begin{equation}
b_n^{+}=-b_{-n},\qquad \hat{P}^{+}=-\hat{P}.
\end{equation}
As we mentioned above, two different bosonizations can be used in constructing 
chain vectors~\eqref{mixed-chain-vector}.  
Moreover, both sets of Feigin-Fuks generators $b_k$ and $b_k^{\text{R}}$ turn out to be relevant.
For example, if one of Young tableaux $Y_2$ or $Y_1$ is empty, then, using $a_k$ and $b_k$ generators,
one finds
\begin{equation}
|Y,\varnothing \rangle=\hat{J}^{-1/b^2}_{Y}(a_{-1}+b_{-1},a_{-2}+b_{-2},\dots)|P\rangle,
\end{equation}
where the polynomials $\hat{J}^{\alpha}_{Y}$ are related with the Jack
polynomials (see Appendix B) as follows
\begin{equation}
\hat{J}^{\alpha}_{Y}(p_1,p_2,\dots)=J^{\alpha}_{Y}(x_1,x_2,\dots)
\end{equation}
and $p_k=( \alpha )^{1/2}(x_1^k+\dots+x_{|Y|}^k)$. In the same time, vectors $|\varnothing,Y\rangle$ 
have rather complicated form in terms of $a_k$, $b_k$. While the second 
bosonization~\eqref{second-bosonization} gives the following simple expression
\begin{equation}
|\varnothing,Y\rangle=
\hat{J}^{-1/b^2}_{Y}(a_{-1}+b_{-1}^{\text{R}},a_{-2}+b_{-2}^{\text{R}},\dots)|P\rangle.
\end{equation}

\section{The orthogonal basis for the case $c=1$}

As we have seen in the previous section the basis vectors have rather simple form if one of Young 
tableaux is empty. But the general answer, when both Young tableaux are non-trivial, remains uncertain.
It turns out that the structure of the orthogonal basis can be clarified completely 
in the case of central charge $c=1$.
The main problem encountered in the previous
analysis is that in general, 
one have to deal with the two non-commuting sets of generators $a_k+b_k$ and $a_k+b_k^{\text{R}}$.
The orthogonality of the basis vectors with one empty diagram  is obvious, given that they correspond to the
 Jack symmetric polynomials, depending on either one, or another set of generators. But the naive conjecture that 
for both non empty Young tableaux
the vectors 
$|Y_1,Y_2\rangle$  are proportional to the product of two Jack 
polynomials, where the first polynomial
depends on $a_k+b_k$ and the second one depends on $a_k+b_k^{\text{R}}$, can not be correct
because of the above mentioned non-commutativity.

For $c=1$ the situation becomes more simple. The reflection relation reduces to the flip of sign 
$b_k^{\text{R}}=-b_k$. Therefore the commutativity between Jack polynomials, corresponding to 
two different 
camps of generators $a_k+b_k$ and $a_k-b_k$ just follows
from the fact that 
\begin{equation}
[a_k+b_k,a_k-b_k]=0. 
\end{equation}
Thus, there exists a natural orthogonal basis
in $\mathcal{A}$ in this case\footnote{Here and below we use short 
notation $\hat{J}_{Y}$ for $\hat{J}_{Y}^{(\alpha=1)}$.}
\begin{equation}
|Y_1,Y_2\rangle=\hat{J}_{Y_1}(a_k+b_k)\hat{J}_{Y_2}(a_k-b_k)|P\rangle
\label{basis-vectors}
\end{equation}
Now we are going to verify that the expansion of the chain vectors 
$|N\rangle$~\eqref{mixed-chain-vector}
in this basis leads
to the AGT representation of the conformal block for $c=1$.
Using orthogonality the coefficients of the vector $|N\rangle$ in the basis $\{|Y_1,Y_2\rangle\}$
\begin{equation}
|N\rangle=\sum_{Y_1,Y_2} C_N(Y_1,Y_2)|Y_1,Y_2\rangle
\label{Decomposition}
\end{equation}
are given in the form
\begin{equation}
 C_N(Y_1,Y_2)=\frac{\langle Y_1,Y_2|N\rangle}{\langle Y_1,Y_2|Y_1,Y_2\rangle}.
\end{equation}
Therefore one have to verify that
\begin{equation}
 C_N(Y_1,Y_2)=\frac{Z_f(\vec{a},\vec{Y}\mu_1),Z_f(\vec{a},\vec{Y},\mu_2)}{D(\vec{a},\vec{Y})},
\label{Coefficients}
\end{equation}
where the parameters are defined in~\eqref{parameters-mu} and~\eqref{parameter-a}.
\paragraph{Level 1:} At level one, the chain vector is
\begin{equation}
|1\rangle^{\text{\sf V}\otimes\mathcal{H}}=|1\rangle^{\text{\sf V}}+|1\rangle^{\mathcal{H}}
=\beta^{\text{\sf V}}_1 L_{-1} |P\rangle+\beta^{\mathcal{H}}_1 a_{-1} |P\rangle,
\end{equation}
where the coefficients are easily found from~\eqref{Vir-chain} and~\eqref{H-chain}
\begin{equation}
\beta^{\text{\sf V}}_1=\frac{\Delta+\Delta_1-\Delta_2}{2\Delta},\qquad \beta^{\mathcal{H}}_1=i \lambda_1.
\end{equation}
The basis vectors are expressed in terms of Jack/Schur symmetric polynomials
\begin{equation}
\begin{aligned}
|\{1\},\varnothing \rangle=
 \hat{J}_{ \{1\}}^{\scriptscriptstyle{(1)}}(a_k-b_k) \hat{J}_{\varnothing}^{\scriptscriptstyle{(1)}}(a_k+b_k)| P\rangle=
 (a_{-1}-b_{-1})|P\rangle,\\
|\varnothing,\{1\} \rangle=
 \hat{J}_{\varnothing}^{\scriptscriptstyle{(1)}}(a_k-b_k) \hat{J}_{ \{1\}}^{\scriptscriptstyle{(1)}}(a_k+b_k)| P\rangle=
 (a_{-1}+b_{-1})|P\rangle,
\end{aligned}
\end{equation}
Using Feigin-Fucks bosonization the first few generators are
\begin{equation}
\begin{aligned}
&L_{-1}=2 i \hat{P} b_{-1} +2 b_{-2} b_1+2 b_{-3} b_2+\dots\,,\\
&L_{-2}=2 i \hat{P} b_{-2} + b_{-1}^2 + 2 b_{-3} b_1+\dots\,,\\
&L_{-3}=2 i \hat{P} b_{-3} +2 b_{-1} b_{-2}+2 b_{-3} b_1+\dots\,.
\end{aligned}
\end{equation}
One can find the scalar products of the chain vector with the basis vectors 
on the level one 
\begin{equation}
\begin{aligned}
& \langle \{1\},\varnothing |1\rangle^{\text{\sf V}\otimes\mathcal{H}}=
-\frac{i(P-\lambda_1-\lambda_2)(P-\lambda_1+\lambda_2)}{2P},\\
& \langle \varnothing, \{1\}|1\rangle^{\text{\sf V}\otimes\mathcal{H}}=
\frac{i(P+\lambda_1-\lambda_2)(P+\lambda_1+\lambda_2)}{2P},
\end{aligned}
\end{equation}
This coincides with the r.h.s. of~\eqref{Coefficients} which looks in this case as follows
\begin{equation}
\frac{Z_f(\vec{a},\vec{Y},\mu_1) Z_f(\vec{a},\vec{Y},\mu_2)}{D(\vec{a},\vec{Y})}=
\frac{(\mu_1-a)(\mu_2-a)}{2iP}.
\end{equation}
\paragraph{Level 2:}
The chain vector on the second level has the form
\begin{equation}
|2\rangle^{\text{\sf V}\otimes\mathcal{H}}=|2\rangle^{\text{\sf V}}+|1\rangle^{\text{\sf V}}|1\rangle^{\mathcal{H}}+|2\rangle^{\mathcal{H}}
=\bigg[\beta^{\text{\sf V}}_{2} L_{-2}+\beta^{\text{\sf V}}_{1,1} L_{-1}^2+
\beta^{\mathcal{H}}_2 a_{-2}+\beta^{\mathcal{H}}_{1,1} a_{-1}^2
+\beta^{\text{\sf V}}_{1} \beta^{\mathcal{H}}_{1} L_{-1} a_{-1}\bigg]|P\rangle,
\end{equation}
where the explicit formulas for the coefficients on the second level:
\begin{equation}
\begin{aligned}
&\beta^{\text{\sf V}}_{2} =-\frac{(\Delta - \Delta^2 - \Delta_1 -
2 \Delta \Delta_1 + 3 \Delta_1^2 - \Delta_2 - 2 \Delta \Delta_2 -
    6 \Delta_1 \Delta_2 + 3 \Delta_2^2)}{(c -
   10 \Delta + 2 c \Delta + 16 \Delta^2)},\\
&\beta^{\text{\sf V}}_{1,1} =\frac{ \Delta + 2 \Delta_1 -  \Delta_2}{
  6 \Delta} + \frac{(c +
      8 \Delta) (\Delta - \Delta^2 - \Delta_1 - 2 \Delta \Delta_1 +
      3 \Delta_1^2 - \Delta_2 -
      2 \Delta \Delta_2 -
      6 \Delta_1 \Delta_2 +
      3 \Delta_2^2)}{12 \Delta (c -
      10 \Delta + 2 c \Delta +
      16 \Delta^2)},\\
&\beta^{\mathcal{H}}_{2}=i \lambda_1,\\
&\beta^{\mathcal{H}}_{1,1}=-2 \lambda_1^2.
\end{aligned}
\end{equation}
$N=2$  basis vectors are
\begin{equation}
\begin{aligned}
&|\{2\},\varnothing \rangle=
 \hat{J}_{ \{2\}}(a_k-b_k) \hat{J}_{\varnothing}(a_k+b_k)|P\rangle=
[ (a_{-2}-b_{-2})+(a_{-1}-b_{-1})^2]|P\rangle,\\
&| \{1,1\},\varnothing \rangle=
 \hat{J}_{ \{1,1\}}(a_k-b_k) \hat{J}_{\varnothing}(a_k+b_k)|P\rangle=
[ (a_{-2}-b_{-2})-(a_{-1}-b_{-1})^2]|P\rangle,\\
&|\varnothing \{2\}\rangle=
\hat{J}_{\varnothing}(a_k-b_k)
\hat{J}_{ \{2\}}(a_k+b_k)|P\rangle=
[ (a_{-2}+b_{-2})+(a_{-1}+b_{-1})^2]|P\rangle,\\
&|\varnothing \{1,1\}\rangle=
\hat{J}_{\varnothing}(a_k-b_k)
\hat{J}_{ \{1,1\}}(a_k+b_k)|P\rangle=
[ (a_{-2}+b_{-2})-(a_{-1}+b_{-1})^2]|P\rangle,\\
& |\{1\},\{1\}\rangle=
 \hat{J}_{\{1\}}(a_k-b_k)
 \hat{J}_{ \{1\}}(a_k+b_k)
 |P \rangle=
 (a_{-1}^2-b_{-1}^2)|P\rangle.
\end{aligned}
\end{equation}
Now the scalar products on the level two can be easily calculated
\begin{equation}
\begin{aligned}
& \langle \{2\},\varnothing |2\rangle^{\text{\sf V}\otimes\mathcal{H}}=
-\frac{(P - \lambda_1 - \lambda_2) (i +
    P - \lambda_1 - \lambda_2) (P - \lambda_1 + \lambda_2) (i +
    P - \lambda_1 + \lambda_2)}{2 P (i + 2 P)},\\
& \langle \{1,1\},\varnothing |2\rangle^{\text{\sf V}\otimes\mathcal{H}}=
-\frac{((P - \lambda_1 - \lambda_2) (-i +
    P - \lambda_1 - \lambda_2) (P - \lambda_1 + \lambda_2) (-i +
    P - \lambda_1 + \lambda_2)}{2 P (-i + 2 P)},\\
& \langle \varnothing, \{2\}|2\rangle^{\text{\sf V}\otimes\mathcal{H}}=
-\frac{(P + \lambda_1 - \lambda_2) (-i +
    P + \lambda_1 - \lambda_2) (P + \lambda_1 + \lambda_2) (-i +
    P + \lambda_1 + \lambda_2)}{2 P (-i + 2 P)},\\
& \langle \varnothing, \{1,1\}|2\rangle^{\text{\sf V}\otimes\mathcal{H}}=
\frac{(P + \lambda_1 - \lambda_2) (i +
   P + \lambda_1 - \lambda_2) (P + \lambda_1 + \lambda_2) (i +
   P + \lambda_1 + \lambda_2)}{2 P (i + 2 P)},\\
& \langle \{1\}, \{1\}|2\rangle^{\text{\sf V}\otimes\mathcal{H}}=
\frac{(P - \lambda_1 - \lambda_2) (P + \lambda_1 - \lambda_2) (P - \
\lambda_1 + \lambda_2) (P + \lambda_1 + \lambda_2)}{1 + 4 P^2}.
\end{aligned}
\end{equation}
Again we can check that this coincides with r.h.s. of~\eqref{Coefficients},
as it follows from the explicit expressions of $Z_f(a,Y,\mu_1)$, $Z_f(a,Y,\mu_2)$ and $D(a)$
for this case given in Appendix A. The details of the similar computations
for the third level are collected in Appendix C. All these checks confirm our conjecture about the decomposition 
of the chain vectors~\eqref{Decomposition} with the Necrasov's coefficients~\eqref{Coefficients} 
in the orthogonal basis~\eqref{basis-vectors}.

\section{Conclusions}
The vectors of the orthogonal basis are expected \cite{F,ALTF} to be
eigenstates of some commutative subalgebra of the universal enveloping algebra
of $\mathcal{A}=Vir \otimes \mathcal H$. (The elements of this
subalgera are called the Integrals of motion.)
First few Integral of motion have been found in~\cite{ALTF} explicitly.
It is not difficult to check  that in the case $c=1$ (or $Q=0$) each of them
can be transformed to the sum of two terms. The first one 
depends only on $a_k+b_k$ and the second one depends on $a_k-b_k$. These terms
are nothing but the Integrals of motion of the two Calogero models expressed in terms
of Heisenberg generators. As it is known, the common eigenstates of the Integrals of
motion of the Calogero model are the Jack polynomials.
Therefore this fact is an additional confirmation of our conjecture
about the factorization of the vectors
 $|Y_1,Y_2\rangle$ in the case $c=1$ into the product of two Jack polynomials.
In general case, i.e. if the central charge $c$ is not equal $1$, the explicit form
of the vectors $|Y_1,Y_2\rangle$ is still an open problem.

\section*{Acknowledgments}
We thank Misha Bershtein, Borya Feigin and Lesha Litvinov for interesting
discussions. The research was held within the framework of the Federal program ``Scientific and Scientific-Pedagogical Personnel of Innovational Russia'' on 2009-2013 (state contracts No. P1339 and No. 02.740.11.5165) and was supported by cooperative CNRS-RFBR  grant PICS-09-02-93064, and by Russian Ministry of Science and Technology under the Scientific Schools grant 6501.2010.2.
\Appendix
\section {AGT relations and Nekrasov partition functions}\label{AGT}
Let $Y=(\lambda_1\geq\lambda_2\geq\dots)$ be a Young tableau, where
$\lambda_i$ is the height of $i$-the column and is equal zero if $i$
is larger than the width of the tableau. Let
$Y^T=(\lambda_1^{'}\geq\lambda_2^{'}\geq\cdots)$ be its transposed.
For a box $s$ at the coordinate $(i,j)$ a function $\phi(a,s)$ and 
its arm-length $A_Y(s)$ and leg-length
$L_Y(s)$ (see figure 1) with respect to the tableau $Y$ are defined as
\begin{equation}
\phi(a,s)=a+b(i-1)+b^{-1}(j-1),
\end{equation}
\begin{equation}
A_Y(s)=\lambda_i - j, \qquad  L_Y(s)=\lambda'_j - i.
\end{equation}
If the box $s$ is outside the tableau, $A_Y(s)$ and $L_Y(s)$ are negative.\\
In the example on the Figure 1:
\begin{equation}
\begin{aligned}
 A_Y(s) &= \text{number of unfilled circles},\\
 L_Y(s) &= \text{number of filled circles}.
\end{aligned}
\end{equation}
\begin{equation*}
  \begin{picture}(30,125)(50,20)
    \Thicklines
    \unitlength 2.3pt 
    \put(20,20){\line(0,1){40}}
    \put(20,40){\line(1,0){30}}
    \put(30,20){\line(0,1){40}}
    \put(20,20){\line(1,0){50}}
    \put(20,50){\line(1,0){20}}
    \put(20,60){\line(1,0){10}}
    \put(20,30){\line(1,0){50}}
    \put(40,20){\line(0,1){30}}
    \put(50,20){\line(0,1){20}}
    \put(60,20){\line(0,1){10}}
    \put(70,20){\line(0,1){10}}
    \put(35,35){\circle*{2}}
    \put(35,45){\circle*{2}}
    \put(34,24){\mbox{$\mathbf{s}$}}
    \put(45,25){\circle{2}}
    \put(55,25){\circle{2}}
    \put(65,25){\circle{2}}
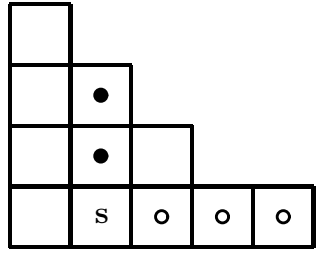
    \put(-20,10){Figure 1: Young diagram, $A_Y(s)$ and $L_Y(s)$.}
    \end{picture}
\end{equation*}
For two Young tableaux $\vec{Y}=(Y_1,Y_2)$ and the vector $\vec{a}=(a_1,a_2)$ 
the function $E$ is defined as
\begin{equation}
E\bigl(a_i-a_j,Y_i,Y_j\bigl|s\bigr)=
(a_i-a_j)-b A_{\scriptscriptstyle{Y_j}}(s)+b^{-1}(L_{\scriptscriptstyle{Y_i}}(s)+1)
\end{equation}
The explicit form of the functions $Z_{\text{\sf{f}}}(\vec{a},\vec{Y},\mu)$,
$Z_{\text{\sf{af}}}(\vec{a},\vec{Y})$ and $Z_{\text{\sf{vec}}}(\vec{a},\vec{Y},\mu)$ are 
\begin{equation}
\begin{aligned}
Z_{\text{\sf{f}}}(\vec{a},\vec{Y},\mu)=\prod_{i=1}^2\prod_{s \in Y_i}(\phi(a_i,s)-\mu+Q),
\end{aligned}
\end{equation}
\begin{equation}
\begin{aligned}
Z_{\text{\sf{af}}}(\vec{a},\vec{Y},\mu)=\prod_{i=1}^2\prod_{s \in Y_i}(\phi(a_i,s)+\mu)
\end{aligned}
\end{equation}
and
\begin{equation}\label{ZfZvec}
\begin{aligned}
Z_{\text{\sf{vec}}}(\vec{a},\vec{Y})=D(\vec{a},\vec{Y})\bar{D}(\vec{a},\vec{Y}),
\end{aligned}
\end{equation}
where the auxiliary functions
\begin{equation}
\begin{aligned}
&D(\vec{a},\vec{Y})=\prod_{i,j=1}^{2}
    \prod_{s\in Y_{i}}E\bigl(a_{i}-a_{j},Y_i,Y_j\bigl|s\bigr),\\
&\bar{D}(\vec{a},\vec{Y})=\prod_{i,j=1}^{2}
    \prod_{s\in Y_{i}}(Q-E\bigl(a_{i}-a_{j},Y_i,Y_j\bigl|s\bigr)).
\end{aligned}
\end{equation}

\section{Jack polynomials}\label{Jack's}
In this appendix we briefly review the main properties of Jack symmetric functions. For more details see
for example \cite{Jack}.

A partition of some natural $n$ is a nonincreasing finite sequence of nonnegative integers
$\lambda = (\lambda_1,\lambda_2,\lambda_3,\dots)$ such that the module $|\lambda| = \sum_i
\lambda_i=n$. Another possible representation of the partition $\lambda = (1^{m_1}2^{m_2}\dots)$,
where $m_k$ is a number of $\lambda_i = k$. Number of nonzero entries in
$\lambda$ is called length of $\lambda$ and denoted by
$l(\lambda)$.
A partition $\lambda = (\lambda_1,\lambda_2,\dots)$ corresponds to a 
Young diagram such that the number of boxes in the $i$th column is
$\lambda_i$.
For two partitions $\lambda$ and $\mu$, if $|\lambda| = |\mu|$ and
\begin{equation}
  \lambda_1+\lambda_2+\cdots+\lambda_i \ge
   \mu_1+\mu_2+\cdots+\mu_i \qquad\text{for all $i$}.
\label{eq:dom}\end{equation}
one says that $\lambda\ge\mu$. This defines a dominance order on the set of partitions.
Let $\lambda$ be a partition with $l(\lambda)\le N$. The monomial symmetric polynomial 
$m_\lambda$ is defined as follows
\begin{equation}
  m_\lambda(x_1,\dots,x_N) =
        \sum_{\sigma(\lambda_i)}
        x_1^{\lambda_{\sigma(1)}}\cdots x_N^{\lambda_{\sigma(N)}},
\end{equation}
where $\alpha = (\alpha_1,\dots,\alpha_N)$ runs over all distinct
permutations of $(\lambda_1,\lambda_2, \dots, \lambda_N)$.
The monomial symmetric functions form a basis in the space of symmetric functions.

The $n$th power sums are related to the monomial polynomials $m_{(n)}$ as
\begin{equation}
  p_n = \sum x_i^n = m_{(n)}.
\end{equation}
For a partition $\lambda = (\lambda_1,\lambda_2,\dots)$, one can define 
$p_\lambda = p_{\lambda_1}p_{\lambda_2}\cdots$.
Then $\{ p_{\lambda} \}$ also form a basis in the space of symmetric functions.
The inner product in $\{ p_{\lambda} \}$ can be defined as
\begin{equation}
  \langle p_\lambda, p_\mu\rangle
  = \alpha^{l(\lambda)} \prod k^{m_k} m_k!\delta_{\lambda\mu},
\end{equation}
where $\lambda =
(1^{m_1}2^{m_2}\cdots)$ and $\alpha$ is some positive real number.

For each partition $\lambda$, there is a unique symmetric polynomial
  $P_{\lambda}^{\alpha}$ satisfying
\begin{equation}
\begin{aligned}
&P_{\lambda}^{\alpha} = m_\lambda + \sum_{\mu < \lambda}
  u_{\lambda\mu}^{(\alpha)}m_\mu, \\
&\langle P_{\lambda}^{\alpha}, P_{\mu}^{\alpha}\rangle = 0 \qquad\text{if}\qquad \lambda\ne\mu,
\end{aligned}
\end{equation}
where $u_{\lambda\mu}^{(\alpha)}$ are some appropriate coefficients. The polynomials
$P_{\lambda}^{\alpha}$ are called the Jack symmetric polynomials. The uniqueness follows from the 
explicit construction of $\{P_{\lambda}^{\alpha}\}$ by means 
of orthogonalization procedure. The nontrivial point here is that the summation runs over $\mu$ 
which is smaller than $\lambda$ with respect to the dominance order.

In a different (so called ``integral'') normalization the Jack polynomials are denoted  
as $J_\lambda^{(\alpha)}$: 
\begin{equation}
  J_\lambda^{(\alpha)} = c_\lambda(\alpha)P_{\lambda}^{\alpha}, \qquad
  c_\lambda(\alpha) = \prod_{s\in Y} (\alpha A_Y(s) + L_Y(s)+1),
\label{eq:normal}\end{equation}
where ``$s\in Y$'' means that $s$ is a box of the Young tableau $Y$
corresponding to $\lambda$ and for ``arm''  and ``leg'' definitions see Appendix A.

We give the explicit form of Jack symmetric functions up to degree 3:
\begin{equation}
\begin{aligned}
 &J_{ \{1\}}^{\scriptscriptstyle{(\alpha)}}=p_1,\\
 &J_{ \{2\}}^{\scriptscriptstyle{(\alpha)}}=\alpha p_2+p_1^2,\\
 &J_{ \{1,1\}}^{\scriptscriptstyle{(\alpha)}}=-p_2+p_1^2,\\
 &J_{ \{3\}}^{\scriptscriptstyle{(\alpha)}}=2\alpha^2 p_3+3\alpha p_2 p_1+p_1^3,\\
 &J_{ \{1,2\}}^{\scriptscriptstyle{(\alpha)}}=-\alpha p_3+(\alpha-1) p_2 p_1+p_1^3,\\
 &J_{ \{1,1,1\}}^{\scriptscriptstyle{(\alpha)}}=2 p_3-3p_2 p_1+p_1^3.
\end{aligned}
\end{equation}

\section{Level 3 computations}
On the third level the mixed chain vector is
\begin{equation}
\begin{aligned}
&|3\rangle^{\text{\sf V}\otimes\mathcal{H}}=|3\rangle^{\text{\sf V}}+|2\rangle^{\text{\sf V}}|1\rangle^{\mathcal{H}}+|1\rangle^{\text{\sf V}}|2\rangle^{\mathcal{H}}+|3\rangle^{\mathcal{H}}=\\
&\bigg[\beta^{\text{\sf V}}_{3} L_{-3}+\beta^{\text{\sf V}}_{1,2} L_{-1}L_{-2}+
\beta^{\text{\sf V}}_{1,1,1} L_{-1}^3+
+\beta^{\mathcal{H}}_3 a_{-3}+\beta^{\mathcal{H}}_{1,2} a_{-1}a_{-2}+\beta^{\mathcal{H}}_{1,1,1} a_{-1}^3+\\
&+\beta^{\text{\sf V}}_{2} \beta^{\mathcal{H}}_{1} L_{-2} a_{-1}+
\beta^{\text{\sf V}}_{1} \beta^{\mathcal{H}}_{2} L_{-1} a_{-2}\bigg]|P\rangle.
\end{aligned}
\end{equation}
The coefficients of the Virasoro chain vector of the 3d level:
\begin{equation}
\begin{aligned}
&\beta^{\text{\sf V}}_{3} =\frac{\Delta \Delta_1 - \Delta_1^2 - \Delta \Delta_1^2 + \Delta_1^3 -
\Delta \Delta_2 -
 3 \Delta_1^2 \Delta_2 + \Delta_2^2 + \Delta \Delta_2^2 +
 3 \Delta_1 \Delta_2^2 - \Delta_2^3}{2 \Delta (2 + c - 7 \Delta + c \Delta +
   3 \Delta^2)},\\
&\beta^{\text{\sf V}}_{1,2} =\frac{12 \Delta^2 - 12 \Delta^3 -
 12 \Delta \Delta_1 -
 36 \Delta^2 \Delta_1 +
 48 \Delta \Delta_1^2 -
 12 \Delta \Delta_2 -
 12 \Delta^2 \Delta_2 -
 72 \Delta \Delta_1 \Delta_2 +
 24 \Delta \Delta_2^2}{-24 c \Delta + 240 \Delta^2 -
 48 c \Delta^2 - 384 \Delta^3}\\
&+\frac{(-24 c \Delta + 168 \Delta^2 -
 72 c \Delta^2 - 216 \Delta^3)}{2 \Delta (2 + c - 7 \Delta + c \Delta +
   3 \Delta^2)}\times\\
&\times\frac{(-\Delta \Delta_1 + \Delta_1^2 + \
\Delta \Delta_1^2 - \Delta_1^3 + \
\Delta \Delta_2 +
  3 \Delta_1^2 \Delta_2 - \Delta_2^2 - \
\Delta \Delta_2^2 -
  3 \Delta_1 \Delta_2^2 + \Delta_2^3)}{(-24 c \Delta + 240 \Delta^2 -
  48 c \Delta^2 - 384 \Delta^3)},\\
&\beta^{\text{\sf V}}_{1,1,1} =
\frac{\Delta + 3 \Delta_1 - \Delta_2}{
  24 \Delta} +\frac{c+8\Delta}{12 \Delta}\times\\
&\times \bigg[\frac{-12 \Delta^2 +
     12 \Delta^3 + 12 \Delta \Delta_1 +
     36 \Delta^2 \Delta_1 -
     48 \Delta \Delta_1^2 +
     12 \Delta \Delta_2 +
     12 \Delta^2 \Delta_2 +
     72 \Delta \Delta_1 \Delta_2 -
     24 \Delta \Delta_2^2}{-24 c \Delta +
     240 \Delta^2 - 48 c \Delta^2 -
     384 \Delta^3} -\\
&- \frac{(-24 c \Delta +
         168 \Delta^2 - 72 c \Delta^2 -
         216 \Delta^3)}{2 \Delta (2 + c - 7 \Delta +
         c \Delta +
         3 \Delta^2)} \times\\
&\times\frac{(-\Delta \Delta_1 + \
\Delta_1^2 + \Delta \Delta_1^2 - \
\Delta_1^3 + \Delta \Delta_2 +
         3 \Delta_1^2 \Delta_2 - \Delta_2^2 - \
\Delta \Delta_2^2 -
         3 \Delta_1 \Delta_2^2 + \
\Delta_2^3)}{ (-24 c \Delta +
         240 \Delta^2 - 48 c \Delta^2 -
         384 \Delta^3)}\bigg]+ \\
&+\frac{(c +
    3 \Delta) (-\Delta \Delta_1 + \
\Delta_1^2 + \Delta \Delta_1^2 - \
\Delta_1^3 + \Delta \Delta_2 +
    3 \Delta_1^2 \Delta_2 - \Delta_2^2 - \
\Delta \Delta_2^2 -
    3 \Delta_1 \Delta_2^2 + \Delta_2^3)}{
 24 \Delta^2 (2 + c - 7 \Delta +
    c \Delta + 3 \Delta^2)} .
\end{aligned}
\end{equation}
The coefficients of the Heisenberg chain vector of the 3d level:
\begin{equation}
\begin{aligned}
&\beta^{\mathcal{H}}_{3}=\frac{2i\lambda_1}{3},\\
&\beta^{\mathcal{H}}_{1,2}=-2 i \lambda_1^2,\\
&\beta^{\mathcal{H}}_{1,2}=-\frac{4i\lambda_1^3}{3}.
\end{aligned}
\end{equation}
On the other side there exist 10  vectors $|Y_1,Y_2\rangle$ of the orthogonal basis
on the third level:
\begin{equation}
\begin{aligned}
&| \{3\},\varnothing\rangle=
 \hat{J}_{ \{3\}}(a_k-b_k) \hat{J}_{\varnothing}(a_k+b_k)|P\rangle=
[ 2(a_3-b_3)+3(a_2-b_2)(a_1-b_1)+(a_1-b_1)^3]|P\rangle,\\
& |\{1,2\},\varnothing\rangle=
 \hat{J}_{ \{1,2\}}(a_k-b_k) J_{\varnothing}(a_k+b_k)|P\rangle=
 [ -(a_3-b_3)+(a_1-b_1)^3]|P\rangle,\\
& |\{1,1,1\},\varnothing\rangle=
 \hat{J}_{ \{1,1,1\}}(a_k-b_k) J_{\varnothing}(a_k+b_k)|P\rangle=
 [ 2(a_3-b_3)-3(a_2-b_2)(a_1-b_1)+(a_1-b_1)^3]|P\rangle,\\
& |\varnothing,\{3\}\rangle=
\hat{J}_{\varnothing}(a_k-b_k)\hat{J}_{ \{3\}}(a_k+b_k)|P\rangle=
[ 2(a_3+b_3)+3(a_2+b_2)(a_1+b_1)+(a_1+b_1)^3]|P\rangle ,\\
&|\varnothing,\{1,2\}\rangle=
 \hat{J}_{\varnothing}(a_k-b_k)\hat{J}_{ \{1,2\}}(a_k+b_k)|P\rangle=
[ -(a_3+b_3)+(a_1+b_1)^3]|P\rangle ,\\
& |\varnothing,\{1,1,1\}\rangle=
 \hat{J}_{\varnothing}(a_k-b_k)\hat{J}_{ \{1,1,1\}}(a_k+b_k)|P\rangle=
 [ 2(a_3+b_3)-3(a_2+b_2)(a_1+b_1)+(a_1+b_1)^3]|P\rangle ,\\
& |\{2\},\{1\}\rangle=
 \hat{J}_{\{2\}}(a_k-b_k)\hat{J}_{ \{1\}}(a_k+b_k)|P\rangle=
 [(a_2-b_2)+(a_1-b_1)^2](a_1+b_1)|P\rangle,\\
& |\{1,1\},\{1\}\rangle=
\hat{J}_{\{1,1\}}(a_k-b_k)\hat{J}_{ \{1\}}(a_k+b_k)|P\rangle=
 [-(a_2-b_2)+(a_1-b_1)^2](a_1+b_1)|P\rangle,\\
& |\{1\},\{2\}\rangle=
 \hat{J}_{\{1\}}(a_k-b_k)\hat{J}_{ \{2\}}(a_k+b_k)|P\rangle=
 (a_1-b_1)[(a_2+b_2)+(a_1+b_1)^2]|P\rangle,\\
& |\{1\},\{1,1\}\rangle=
 \hat{J}_{\{1\}}(a_k-b_k)\hat{J}_{ \{1,1\}}(a_k+b_k)|P\rangle=
(a_1-b_1)[-(a_2+b_2)+(a_1+b_1)^2]|P\rangle.\\
\end{aligned}
\end{equation}
With the matrix of scalar products
$M=\langle a_{k_1}\cdots b_{m_1}\cdots| a_{-l_1}\cdots L_{-n_1}\cdots\rangle$
for $\sum k_i+\sum m_i=\sum l_i+\sum n_i=3$
\begin{equation}
M=
 \begin{pmatrix}
 3 i P & 6 i P & 6 i P  & 0 & 0     & 0 &0 & 0 &0&0\\
 1    & 1-2P^2& -6P^2  & 0 & 0     & 0 &0 & 0 &0&0\\
 0    & \frac{3 i P}{2}& -6i P^3   & 0 & 0 & 0 &0 & 0 &0&0\\
 0    &     0           &     0     & \frac 32 & 0 & 0 &0 & 0 &0&0\\
 0    &     0           &     0     & 0        &\frac 12 & 0 &0 & 0 &0&0\\
0    &     0           &     0     & 0          &0       & \frac 34 & 0 & 0 &0&0\\
0    &     0           &     0     & 0 & 0 & 0 &    i P  & i P&0&0\\
0    &     0           &     0     & 0 & 0 & 0& \frac 14 &-P^2 &0&0\\
0    &     0           &     0     & 0 & 0 & 0 &0 & 0 &i P&0\\
0    &     0           &     0     & 0 & 0 & 0 &0 & 0 &0 &\frac{i P}2
\end{pmatrix}\ ,
\end{equation}
one can find for the scalar products of the vectors $|Y_1,Y_2\rangle$ with the chain vector
$|3\rangle^{\text{\sf V}\otimes\mathcal{H}}$ \\
the following answers:
\begin{eqnarray*}
\begin{aligned}
 \langle \{3\},\varnothing |3\rangle^{\text{\sf V}\otimes\mathcal{H}}=&(4 i P (-i + P) (-i + 2 P))^{-1}
(P + \lambda_1 - \lambda_2) (-i +
   P + \lambda_1 - \lambda_2) \\
&\times(-2 i +
   P + \lambda_1 - \lambda_2)
 (P + \lambda_1 + \lambda_2) (-i +
   P + \lambda_1 + \lambda_2) (-2 i + P + \lambda_1 + \lambda_2),\\
 \langle \{1,2\},\varnothing |3\rangle^{\text{\sf V}\otimes\mathcal{H}}=&(2 i P  (1 + 4 P^2))^{-1}
(P + \lambda_1 - \lambda_2) (P + \lambda_1 + \lambda_2) \\
&\times(1 + P^2 + 2 P \lambda_1 + \lambda_1^2 - 2 P \lambda_2 -
   2 \lambda_1 \lambda_2 + \lambda_2^2) \\
&\times(1 + P^2 +
   2 P \lambda_1 + \lambda_1^2 + 2 P \lambda_2 +
   2 \lambda_1 \lambda_2 + \lambda_2^2),\\
 \langle \{1,1,1\},\varnothing |3\rangle^{\text{\sf V}\otimes\mathcal{H}}=&
(4 (-1 - i P) P (-i + 2 P))^{-1}(P - \lambda_1 - \lambda_2) (-i + P - \lambda_1 - \lambda_2)\\
&\times (-2 i + P - \lambda_1 - \lambda_2) (P - \lambda_1 + \lambda_2) \
(-i + P - \lambda_1 + \lambda_2) (-2 i +
   P - \lambda_1 + \lambda_2),\\
 \langle \varnothing, \{3\}|3\rangle^{\text{\sf V}\otimes\mathcal{H}}=&(-4 i P (i + P) (i + 2 P))^{-1}
(P - \lambda_1 - \lambda_2) (i + P - \lambda_1 - \lambda_2) \\
&\times (2 i +
    P - \lambda_1 - \lambda_2) (P - \lambda_1 + \lambda_2) (i +
   P - \lambda_1 + \lambda_2) (2 i + P - \lambda_1 + \lambda_2),\\
 \langle \varnothing, \{1,2\}|3\rangle^{\text{\sf V}\otimes\mathcal{H}}=&(-2 i P (1 + 4 P^2))^{-1}
(P - \lambda_1 - \lambda_2) (P - \lambda_1 + \lambda_2) \\
&\times(1 +
   P^2 - 2 P \lambda_1 + \lambda_1^2 + 2 P \lambda_2 -
   2 \lambda_1 \lambda_2 + \lambda_2^2) \\
&\times(1 + P^2 -
   2 P \lambda_1 + \lambda_1^2 - 2 P \lambda_2 +
   2 \lambda_1 \lambda_2 + \lambda_2^2),\\
 \langle \varnothing, \{1,1,1\}|3\rangle^{\text{\sf V}\otimes\mathcal{H}}=&
(4 (-1 - i P) P (-i + 2 P))^{-1}(P - \lambda_1 - \lambda_2) (-i +
   P - \lambda_1 - \lambda_2) \\
&\times (-2 i +
   P - \lambda_1 - \lambda_2) (P - \lambda_1 + \lambda_2) (-i +
   P - \lambda_1 + \lambda_2) (-2 i + P - \lambda_1 + \lambda_2),\\
 \langle \{2\}, \{1\}|3\rangle^{\text{\sf V}\otimes\mathcal{H}}=&(4 (-1 - i P) P (i + 2 P))^{-1}
(P - \lambda_1 - \lambda_2) (P + \lambda_1 - \lambda_2) \\
&\times(-i +
   P + \lambda_1 - \lambda_2) (P - \lambda_1 + \lambda_2) (P + \
\lambda_1 + \lambda_2) (-i + P + \lambda_1 + \lambda_2),\\
 \langle \{1,1\}, \{1\}|3\rangle^{\text{\sf V}\otimes\mathcal{H}}=&(4 (-1 - i P) P (i + 2 P))^{-1}
(P - \lambda_1 - \lambda_2) (P + \lambda_1 - \lambda_2) \\
&\times(i +
   P + \lambda_1 - \lambda_2) (P - \lambda_1 + \lambda_2) (P + \
\lambda_1 + \lambda_2) (i + P + \lambda_1 + \lambda_2),\\
 \langle \{1\}, \{2\}|3\rangle^{\text{\sf V}\otimes\mathcal{H}}=&(4 (-1 + i P) P (-i + 2 P))^{-1}
(P - \lambda_1 - \lambda_2) (i+P - \lambda_1 - \lambda_2) \\
&\times(
   P + \lambda_1 - \lambda_2) (P - \lambda_1 + \lambda_2) (i +P -\lambda_1 +
\lambda_2) (P + \lambda_1 + \lambda_2),\\
 \langle \{1\}, \{1,1\}|3\rangle^{\text{\sf V}\otimes\mathcal{H}}=&
(4 (-1 + 2i P) P (-i + P))^{-1}
(P - \lambda_1 - \lambda_2) (-i+P - \lambda_1 - \lambda_2) \\
&\times(
   P + \lambda_1 - \lambda_2) (P - \lambda_1 + \lambda_2) (-i +P -\lambda_1 +
\lambda_2) (P + \lambda_1 + \lambda_2).
\end{aligned}
\end{eqnarray*}
These expressions coincide with the Nekrasov's coefficients~\eqref{Coefficients}.

\providecommand{\href}[2]{#2}\begingroup\raggedright

\end{document}